\documentstyle[aps,twocolumn,prl,epsf]{revtex}
\newcommand{\be}{\begin{equation}}\newcommand{\bea}{\begin{eqnarray}}
\newcommand{\ee}{\end{equation}}\newcommand{\eea}{\end{eqnarray}}

\begin{document}
\title{Giant Charge Inversion of a Macroion Due to Multivalent \\
Counterions and Monovalent Coions: Molecular Dynamics Study}

\author{Motohiko Tanaka$ {}^{1} $, and A.Yu Grosberg$ {}^{2} $}

\address{
$ {}^{1} $National Institute for Fusion Science, Toki 509-5292, Japan \\
$ {}^{2} $Department of Physics, University of Minnesota,
Minneapolis, MN 55455}

\address{\begin{quote}
{\em We report molecular dynamics simulation of the (overall
neutral) system consisting of an immobile macroion surrounded by
the electrolyte of multivalent counterions and monovalent coions.
As expected theoretically, counterions adsorb on the macroion
surface in the amount much exceeding neutralization requirement,
thus effectively inverting the sign of the macroion charge. We
find two conditions necessary for charge inversion, namely,
counterions must be multivalently charged and Coulomb interactions
must be strong enough compared to thermal energy. On the other
hand, coion condensation on the multivalent counterions similar to
Bjerrum pairing is the major factor restricting the amount of
charge inversion.  Depending on parameters, we observe inverted
charge up to about 200\% the original charge of the macroion in
absolute value. The inverted charge scales as $\sim \zeta^{1/2}$
when $\zeta <1$ and crosses over to $\sim \zeta$ for $\zeta > 1$,
where $\zeta = (A_0/r\!_s )^2$, $r\!_s$ is the Debye
screening length in the electrolyte and $ A_0 $ is the distance
between adsorbed counterions under neutralizing conditions.  These
findings are consistent with the theory of "giant charge
inversion" [{\it Phys.Rev.Lett.}, {\bf 85}, 1568 (2000)].
}\end{quote}}

\address{PACS numbers: 61.25.Hq, 61.20.Qg, 87.14.Gg, 82.70Dd}

\maketitle

\section{Introduction}

Correlation effects in the systems of charged particles, such as
plasma or electrolyte solution, are well known since the works by
Debye and H\"{u}ckel in 1923 \cite{DH}.  Classical intuition
suggests that correlation can be viewed as screening in which a
cloud of ions around, say, positive particle is slightly dominated
by negative counterions, such that for an outside observer (who
measures the electric field) the shield of predominantly negative
charges effectively reduces the central positive charge. Recently,
a significant attention has been attracted by the notion that much
more dramatic effect is possible in the system with strongly
charged ions \cite{Shklov}.  Namely, instead of charge reduction
due to the shielding, it is possible to observe charge inversion
due to the "over-screening".  Furthermore, it was shown a year ago
that the inverted charge may be quite large, even larger in
absolute value than the original bare charge, giving rise to the
concept of "giant" charge inversion \cite{NguGS}.

In the present paper, we use molecular dynamics simulation
technique to address the question of possible limits  of charge
inversion.  Overall, we confirm the theoretical prediction
\cite{NguGS} and observe "giant" charge inversion, with the ratio
of inverted and bare charges reaching up to about 1.6 (in absolute
value).

Although we consider here only primitive schematic model with
spherical ions immersed in the medium of a constant dielectric
permeability $\epsilon$, this should be viewed as the step towards
better understanding of such first magnitude scientific problems
as, e.g., that of chromatin structure.  Indeed, chromatin
represents a complex of strongly negatively charged thread of DNA
with positively charged smaller protein molecules.   For instance,
virtually every paper on charge inversion mentions the fact that
protein core of a nucleosome particle \cite{Nature} carries lesser
amount of positive charge than the amount of negative charge on
the wrapped around DNA.  On a simpler level, complexes of
polycations and polyanyons were under scrutiny for a long time
\cite{Kabanov}, as well as complexes of charged polymers with
charged colloids \cite{Dubin}.

In theoretical aspect, the most advanced treatment of charge
inversion is due to Shklovskii and his co-workers
\cite{Shklo1,Shklo2,Shklo3,NguGS}. In these works, the universal
physical mechanism behind charge inversion is recognized as
correlations between shielding ions (see also brief review article
\cite{Shklo4}).  In the work \cite{Shklo1}, the idealized image of
these shielding counterions forming a Wigner crystal on the
surface of the shielded macroion was emphasized (see also earlier
work \cite{RB}).  However, it was mentioned in \cite{Shklo1} and
addressed in more details in \cite{Shklo2,Shklo3} that in most
real cases, correlations are not quite as strong as to produce a
crystal, but sufficient to maintain short range order, and,
therefore, correlation energy is similar to that of a crystal.
Obviously, this mechanism is operational when shielding ions are
strongly charged. Furthermore, it was realized that the best
situation for charge inversion occurs when monovalent salt is
present in addition to strongly charged ions \cite{NguGS} (see
also \cite{Joanny}). Salt ions, as their charges are small, behave
in a "traditional" way; they simply screen all interactions at the
distance about Debye length $r_s$. However trivial itself, this
leads to a dramatic increase of charge inversion, because the
attraction of a counterion to its Wigner-Seitz cell on the
macroion surface is over a significantly shorter range than the
repulsion of a counterion from the uncompensated charge of all
other counterions.  For completeness, we mention here also recent
works developing charge inversion theory to include
polyelectrolyte ions \cite{Shklo5,Shklo6,Rubinstein}, as well as
more formal theoretical approaches \cite{Orland}.

Charge inversion has been seen several times in simulations,
starting from the pioneering work \cite{sim-pio}.  In recent works
\cite{Allah,Matees,Lins,Mainz,mainz2,mainz3,Netz} computer
simulations were reported along with various ways to re-derive and
re-examine the concept of lateral correlations between counterions
as the driving force behind charge inversion.  The authors of
\cite{Netz,Mainz} reported quite impressive agreement between
theoretical conjectures and their computation data.  However
sophisticated, these simulations concentrated on the cases of no
added salt and of abundance of counterions.  In other words, they
only examined the very dilute extreme with respect to macroions
assuming at the same time finite concentration of counterions. Our
first intent in the present work is to relax this serious
restriction and to simulate a realistic model in which
thermodynamic cost of adsorption of counterions on the surface of
a macroion is contributed by both the events on the macroion
surface and in the surrounding solution.

The other closely connected goal of our present study has to do
with the following delicate aspect of the "giant" charge inversion
scenario.  In order to make correlations and charge inversion
stronger, one is tempted to choose larger ratio of Coulomb energy
to thermal energy.  But when it is too large, the small salt ions
start to condense on the surfaces of counterions, effectively
reducing their charge. Therefore, charge inversion is expected to
be the strongest in the intermediate regime, when correlations
between counterions are already strong, but condensation of small
ions on them is still weak.  Therefore, we want to check in the
present work computationally how robust is this theoretical
prediction.

To achieve the above stated goals, we perform in the present paper
molecular dynamics study of the system consisting of a single
macroion, large number of multivalent counterions, and a multitude
of monovalent coions immersed in a Langevin fluid.  It is worth
noting that hydrodynamic effects, which may be of significant
importance for interactions between colloidal particles away from
thermodynamic equilibrium \cite{Crock,Kepl,Michael}, are totally
ignored in the present study, because we concentrate on the
equilibrium aspects only.

The paper is organized as follows. The simulation method and
parameters are described in Section \ref{sec:method}.  In Section
\ref{sec:results}, by direct measurement of the peak height of the
radial charge distribution we show that giant charge inversion
takes place when the following two conditions are simultaneously
met: (1) multivalent counterions with valence $ Z \ge 2 $ are
present, and (2) Coulomb energy prevails over the thermal energy
at the length scale of a single ion size, $a$:  $ \Gamma = Z^{2}
e^{2} / \epsilon a k_{B}T > 1 $.  We study in details the
dependence of charge inversion on the radius and charge of the
macroion, the valence and density of counterions and coions, and
temperature. For large density and valence of counterions, the
amount of inverted charge increases linearly with ionic strength,
and reaches up to 200\% the original macroion charge.

Extension of the present work to the case under electrophoretic
environments is discussed in a separate paper \cite{Tan} in which
the effect of an applied electric field on the charge inversion
process is investigated with the use of molecular dynamics
simulation.

\section{Simulation Method and Parameters}\label{sec:method}

\subsection{Equations}

Specifically, we consider the following model.  The system
includes: a single macroion with negative charge $ Q_{0}  < 0 $, $
N^{+} $ multivalent counterions with a positive charge $ Ze $
each, and $ N^{-} $ monovalent coions with a negative charge $
(-e) $ each ($ e > 0 $ is the elementary charge). Overall charge
neutrality is strictly enforced: $ Q_{0}  + ZeN^{+} - eN^{-} = 0
$. All ions are confined within the three-dimensional simulation
domain having spherical shape with radius $ R_{M} $. The macroion
is considered immobile; it is placed at the origin (center of the
domain), and all other ions are mobile.  All ions are supposed to
be of spherical shapes, with macroion having radius $R_{0} $ and
all mobile ions having identical radius $a$; $a$ serves also as a
unit of length.

The (classical) molecular dynamics simulation solves the
Newton-Langevin equations of motion
\begin{eqnarray}
\label{eqM} m \frac{d{\bf v}_{i}}{dt} & = & - \nabla \Phi({\bf
r}_{i})
 - \nabla \phi({\bf r}_{i})
  - \nu a {\bf v}_{i} + {\bf F}_{th} \ , \nonumber  \\
  \frac{d{\bf r}_{i}}{dt} & = & {\bf v}_{i} \ ,
\end{eqnarray}
where the potentials $\Phi$ and $\phi$, describe interactions of a
given ion with other mobile ions and with the macroion,
respectively:
\begin{eqnarray}
&& \Phi({\bf r}_{i}) = \sum_{j} \left\{
   \frac{Z_{i} Z_{j} e^{2}}{\epsilon r_{ij}}
   + \epsilon_{LJ} \left(
      \left(\frac{a}{r_{ij}}\right)^{12} -
      \left(\frac{a}{r_{ij}}\right)^{6} \right) \right\} \ ;
\nonumber \\
      && \phi ({\bf r}_{i}) = Z_{i}e \ \frac{Q_{0} }{\epsilon r_{i}}  \
.
\end{eqnarray}
Here, $ {\bf r}_{i} $ and $ {\bf v}_{i} $ are the position and
velocity vectors of the $ i $-th particle, $ r_{ij} = |{\bf r}_{i}
-{\bf r}_{j}| $, $ \epsilon $ the dielectric constant, $
\epsilon_{LJ} $ the Lennard-Jones energy.  As regards boundaries,
we assume elastic reflection every time when a mobile ion hits
either the domain boundary at $r=R_{M}$ or the macroion surface at
$ r= R_{0}  $. The last two terms of Eq.(\ref{eqM}) represent the
Langevin thermostat due to surrounding neutral medium. The Stokes
formula for a sphere is adopted for the friction term with $ \nu $
being the friction constant, and $ {\bf F}_{th} $ is the random
$\delta$-correlated thermal agitation.

The inertia term is retained in the momentum equation for
numerical stability of the electrostatic forces, masses of all
mobile ions are assumed identical, equal to $ m $.  This leads to
the choice of $\omega_{p}^{-1} $ as the natural time unit, where $
\omega_{p}= (4 \pi n_0  e^{2}/\epsilon m)^{1/2} $ is plasma
frequency and $ n_0  $ the average ion number density.

\subsection{Parameters}\label{sec:numbers}

It must be born in mind that phenomena resembling charge inversion
may occur when other forces, apart from Coulomb electrostatic
ones, operate in the system (including complicated helical shape
of the molecules involved; see, for instance, \cite{Leikin}). In
this study we are interested in the situation when pure
electrostatic forces dominate.  Accordingly, we choose $
\epsilon_{LJ} = (1/12) e^{2}/\epsilon a $; this corresponds to the
depth of Lennard-Jones potential well equal to $-\epsilon_{LJ}/4 =
- (1/48) e^{2}/\epsilon a$, which means that Lennard-Jones
attraction force is very small compared to Coulomb force even at
the distance of ion size $a$ and even for monovalent ions.

We also consider densities at which short range repulsion
(excluded volume effect) between ions is not important, as volume
fraction of particles in the simulation domain, $ \phi = \phi^{+}
+ \phi^{-} = a^3 (N^{+} + N^{-})/(R_{M}^3 - R_{0} ^3)$, is small,
about $\phi \approx 0.05$ or less for all cases considered in this
paper.

By contrast, Coulomb interactions are strong.  To be more
specific, there are several relevant parameters controlling
different manifestations of Coulomb forces.  First of all,
multivalent $Z$-ions are attracted to the macroion and can be
adsorbed on its surface.  This is controlled by the parameter \be
\Gamma_{Q} = \frac{ZeQ }{\epsilon R_{0}  T} \label{eq:GammaQ} \ee
(for the temperature $T$, we use energy units and omit
Boltzmann constant $k_B$). Second, monovalent coions are attracted
to the multivalent counterions ($Z$-ions), and can condense there,
which is controlled by the parameter \be \Gamma_a = \frac{ Z e^2}{
\epsilon a T} \ . \label{Eq:Gammaa} \ee A little more delicate
matter is the possible correlation between repelling ions,
particularly those adsorbed on the macroion.  This is
characterized by $ \Gamma = Z^{2} e^{2}/\epsilon A T $, where $A$
can be estimated as the distance between two adsorbed counterions
in the situation when the number of adsorbed counterions is just
sufficient to neutralize the macroion, that is $(Q/Z e) \pi
(A/2)^2 = 4 \pi R_{0} ^2 $ or $A = 4 R_{0}  \sqrt{Ze/Q}$.  Thus,
\be \Gamma = \frac{ Z^{3/2} e^{3/2} Q^{1/2} }{ 4 \epsilon R_{0}
T} \ . \label{eq:Gamma} \ee In principle, there is also other
similar $\Gamma$ parameters which control correlations between
various ions in the bulk; in this work we do not address this
aspect.

In the present study, we typically look at the $\Gamma_a$ values
in the range $ \Gamma = 6 \sim 80 $.   For the estimates, it is
useful to keep in mind that Bjerrum length $\ell_B = e^2/\epsilon
T$ is close to 7 \AA under typical conditions - at room temperature
in water ($\epsilon \approx 80$).  In particular, for the typical
small ions, for which $a \approx 4 {\rm \AA}$ (counting attached
water), we get $\Gamma_a \approx 1.7 Z {\rm \AA}$, which is
roughly between 4 and 10 for $Z$ between 2 and 7.  As regards
$\Gamma = \Gamma_a \left( a / 4 R_{0}  \right) \sqrt{ ZQ/e}$, it
may be greater than $\Gamma_a$ if macroion is
strongly charged ($Q/e$ is large).

Note also that under typical conditions, such as  $m \approx 50
m_{H} $ and $n_0 \approx (1/10 {\rm \AA})^3$,  where $m_{H}$ is
proton mass, and $ n_0  $ the average density of counterions, the
characteristic frequency and time are about $ \omega_{p} \approx
6.6 \times 10^{11} s^{-1} $ and $ \omega_{p}^{-1} \approx 1.5 ps
$.

In our molecular dynamics experiment, the initial positions of co-
and counterions are distributed randomly between the two spheres $
R_{0}  < r < R_{M} $, each ions having the velocity that satisfies
the Maxwell distribution. The integration of the equations of
motion is done with the use of the leapfrog method which is
equivalent to Verlet algorithm\cite{Verl}.  The time step of
integration is $ \Delta t= 0.01 \omega_{p}^{-1} $, and simulation
runs are executed up to $ 5000 \omega_{p}^{-1} $ at which time the
peak height of the inverted charge Eq.(\ref{eq:int-ch}) has become
stationary.

Below, in Section \ref{sec:results}, we report the simulation
results concentrating on the general properties of the charge
inversion: its dependence on the radius and charge of a macroion,
the valence and density of counterions, and temperature. While
changing the parameters, the electrostatic binding energy of
counterions to the macroion is kept constant by fixing $
\Gamma_{Q}$ Eq.(\ref{eq:GammaQ}).

In the present study, the following values of parameters are
considered "standard" and used unless otherwise specified: radius
of the macroion $ R_{0} = 3a $, its charge $ Q_{0} = -28e $
(assumed negative), valence of the counterions $ Z= 7 $, and the
number of the counterions and coions $ N^{+}= 52 $ and $ N^{-}=
336 $, respectively.  The radius of the outer boundary sphere is $
R_{M}= 20a $.  The temperature is chosen such that $ \Gamma_a =
4.2 Z $.

To support physical intuition, it is useful to estimate the Debye
screening length.  Naive application of standard formula yields
\begin{eqnarray}\label{eq:rs}
r_s & = & \left[ 4 \pi e^2 \frac{ \left. Z \right. ^2 N^{+} +
N^{-} }{\frac{4}{3} \pi (R_{M}^3-R_{0} ^3) \epsilon T}
\right]^{-1/2}  \nonumber \\ & = & a \left[ \frac{(R_{M}/a)^3 -
(R_{0} /a)^3}{3 \Gamma_{a} \left( Z N^{+} + Z^{-1} N^{-}
\right)} \right]^{1/2} \ ,
\end{eqnarray}
which is about $0.5 a$ under the "standard" conditions.  This
result may seem surprising, as physically screening length cannot
be smaller than the size of smallest ions \cite{comment}.  Of
course, such a small value of screening length indicates very
strong Coulomb interactions in the bulk solution.  This fact can
be also seen differently, by noting that the parameters
controlling validity of the linearized Debye-H\"uckel theory for
the plasma away from macroion are $ Z \phi_{+}^{1/3} \Gamma_a$ and
$ \phi_{-}^{1/3} \Gamma_a/Z$, and they are both large compared to
unity, about 10--200 ($ Z $=3--7) and 1--4, respectively. (These
parameters mean ratio of Coulomb between particles of respective
signs and thermal energies at typical distances - controlled by
densities.)
Thus, we consider the conditions under which
plasma outside the macroion is very nonlinear.  Physically, this
is manifested by extensive condensation of coins on multivalent
counterions, as will be seen in the results below.  Such
condensation is analogous to Bjerrum pairs formation. Condensation
means that effective charges of particles are reduced, and also
the effective density of charged particles is lowered.  That leads
to the increase of real screening radius which attains some
respectable value.  We do not attempt to estimate it, as we do not
rely on any particular theory.  Instead, we will just see what
molecular dynamics show. We shall see that condensation of coions is
the major factor limiting the extent of charge inversion.

Another interesting quantity to estimate is Gouy-Chapman length
associated with the surface of a macroion,
\begin{equation}\label{eq:lambda}
\lambda = \frac{\epsilon T}{2 Z \pi e \sigma} = a \frac{2 (R_{0}
/a)^2 }{ \Gamma_{a} \left| Q_{0} /e \right|}
\end{equation}
(where $\sigma = \left| Q_{0}  \right| /4 \pi R_{0} ^2$)  turns
out to be about $0.15 a / Z$.  (Strictly speaking, $\lambda$ is
defined for the plane, not spherical surface; however, since
$\lambda / R_{0}  \approx 0.05 \ll 1$, defining $\lambda$ based on
plane geometry is reasonable.)

For the standard run, it takes about $ 2.5 \times 10^{3}
\omega_{p}^{-1} $ before a state is reached which can be assumed
equilibrated, at least in terms of the inverted charge being
stationary.

\section{Simulation Results}\label{sec:results}

\subsection{Observing charge inversion}

\subsubsection{Standard regime}

\begin{figure} \centerline{\epsfxsize=0.4\textwidth
\epsffile{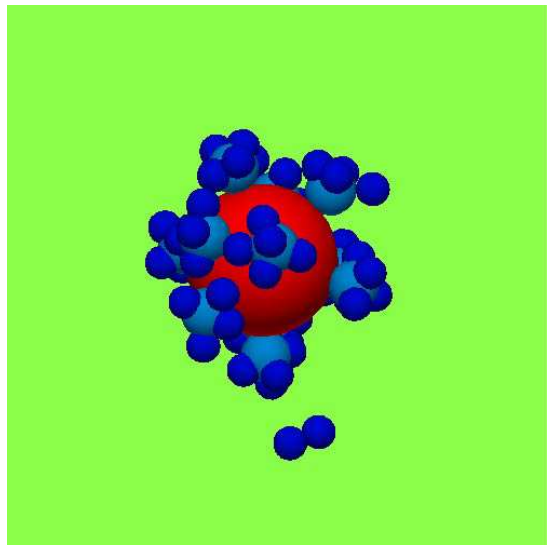} } \vspace*{0.3cm}

%\vspace*{5.7cm}
\caption{The bird's-eye view of the screening atmosphere within
$ 3a $ from the macroion under "standard" conditions.
Macroion is the red ball in the middle.
Multivalent counterions of valence $ Z= 7 $ and monovalent coions
are shown in dark blue and light blue, respectively.  Macroion
radius $ R_{0} = 3a $, charge $ Q_{0} = -28e $.  Temperature is
chosen such that  $ \Gamma_a = 29.4 $.  Note that significant
condensation of coions on the counterions is observed.
For this reason, correlations between adsorbed multivalent
counterions are nowhere near ideal Wigner crystal while
$\Gamma \approx 137 $ (Eq.(\protect\ref{eq:Gamma})) is very
large.}
\label{standard_bird_eye}
\end{figure}

\begin{figure} \centerline{\epsfxsize=0.4\textwidth
\epsffile{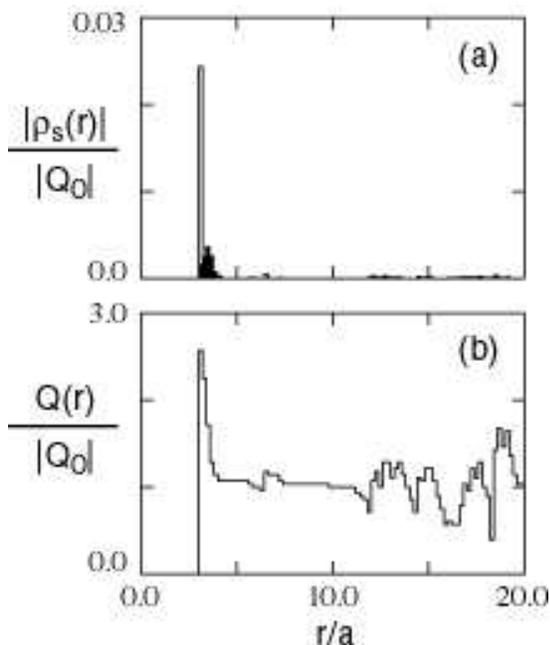} }

\caption{Charge inversion under "standard conditions", as in Figure
\protect\ref{standard_bird_eye}. (a) The radial distribution
function of the charge $ \rho_{s}(r) $
(Eq.(\protect\ref{eq:smallq})) of counterions (open bars) and that
of coions (shaded bars) as a function of the distance $ r $ from
the macroion center. (b) The integrated charge distribution $ Q(r)
$ of counterions plus coions (Eq.(\protect\ref{eq:int-ch})). The
portion $ Q(r)/|Q_{0} | > 1 $ corresponds to charge
inversion.}\label{standard_plots}
\end{figure}

\begin{figure}
\centerline{\epsfxsize=0.4\textwidth \epsffile{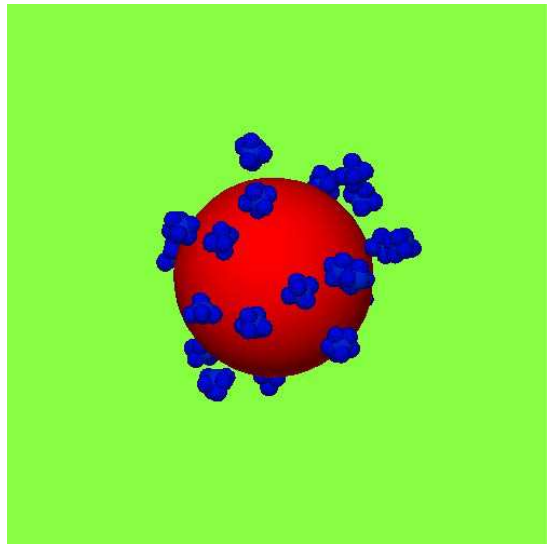} }
\vspace*{0.3cm}

%\vspace*{6.0cm}
\caption{The bird's-eye view of the
screening atmosphere within $ 3a $ from the "large" macroion:
macroion radius $ R_{0} = 8a $, charge $ Q_{0} = -28e $.
Temperature is adjusted such that $ e^{2}/ \epsilon R_{0}  T =
const $ is the same as in Fig.\protect\ref{standard_bird_eye}.
This means that $\Gamma$ Eq.(\protect\ref{eq:Gamma}) is the same
here and in Fig.\protect\ref{standard_bird_eye}, while
$\Gamma_a$ is greater here than in
Fig.\protect\ref{standard_bird_eye} by a factor of
$8/3$.  Accordingly, stronger binding of monovalent coions
(dark blue) to multivalent counterions (light blue) is observed,
and adsorbed counterions are less strongly correlated.}
\label{large_bird_eye}
\end{figure}

The results of our simulations are presented in the Figures
\ref{standard_bird_eye}--\ref{figure8}. Figures
\ref{standard_bird_eye} and \ref{standard_plots} present typical
results of runs performed under what we call "standard"
conditions. Specifically, Fig. \ref{standard_bird_eye} shows a
snapshot of the spatial distribution of counterions and coions
around the macroion after charge distribution has become
stationary.  Since our simulation includes hundreds of particles,
it is impossible to "see" them in any meaningful way; what we can
see, however, is the configuration of ions in the immediate
vicinity of the macroion surface.  This is shown in Fig.
\ref{standard_bird_eye} in which only the ions residing in the
thin layer $ R_{0}  \le r \le R_{0} +3a $ are depicted.

As seen in Fig. \ref{standard_bird_eye}, counterions (light blue)
attach right on the surface of the macroion with a lateral
spacing, while coions (dark blue) stay some distance away from the
macroion surface. It is clear that lateral correlations are
present between counterions, particularly because there are no
pairs in which counterions are close to each other.  Not
surprisingly, however, this correlations are much weaker than in
the case without coions examined in \cite{Mainz}:  although
counterions are correlated in Fig. \ref{standard_bird_eye},
their spacings are not regular and cannot be identified as Wigner
crystal.  As regards coions, they are seen to condense on the top
side of the counterions, presumably because of strong repulsion of
the coions from macroion surface.  We note here that this
condensation of coions on the counterions is responsible for
limiting the amount of charge inversion. In the configuration
shown in Fig. \ref{standard_bird_eye}, the numbers of
counterions and coions within the distance $ a $ from the macroion
surface are $ N^{+}= 11 $ and $ N^{-}= 5 $, respectively. This
means that the net charge of the entire complex, i.e. "macroion
$+$ attached counterion $+$ attached coions", is $ +44e $.  This
is to be compared with the bare macroion charge of $-28e$, which
amounts to charge inversion of about 160\% the original macroion
charge.

\vspace{-.3cm} 

Figure \ref{standard_plots} (a)  shows the radial distributions of
co- and counter-ions charges
\be \label{eq:smallq} \rho_s(r)= e Z_s \int \sum_{i \in s}
\frac{\delta ({\bf r}-{\bf r}_{si})}{ 4 \pi r_{si}^2} d
\Omega_{\bf r}  \ , \ee
where $s$ means either co- or counter-ions, $ Z_s $ is,
accordingly, either $ -1 $ or $ Z $; summation runs over all ions
of the given sort $s$, ${\bf r}_{si}$ is the position vector of
ion $i$ of the sort $s$, and $\Omega_{\bf r}$ is the solid angle
of directions of vector ${\bf r}$. These results are consistent
with the conclusion of 160 \% charge inversion. Indeed, the
distribution of the counterions, denoted by open bars, is sharply
peaked at $ r \cong R_{0}  $, while that of the coions (shaded
bars) is broad and detached from the macroion surface. Although at
this stage we do not formulate any rigorous algorithmic definition
as to which counterions are close enough to the macroion to be
called "bound," we note that the peak in the radial density
distribution of counterions is sharp enough to provide for quite
clear distinction between bound and unbound ions. We therefore
rely on this sharp peak, and in what follows we describe as bound
those counterions which belong to this peak.

\vspace{-.3cm} 

Figure \ref{standard_plots} (b) depicts the integrated charge of
the movable ion species (counterions and coions) of Fig.
\ref{standard_plots} (a), starting at the surface of the macroion,
\begin{eqnarray}
\label{eq:int-ch}
  && Q(r) = \sum_{s} \int_{R_{0} }^{r} \rho_{s}(r^{\prime}) \ 4 \pi
\left. r
^{\prime} \right. ^2 \ d r^{\prime} \ .
\end{eqnarray}
The portion above the baseline $ Q/|Q_{0} |= 1 $ corresponds to
the charge inversion (this applies to all the following figures).
The net amount of inverted charge reaches 160\% for this run, as
stated above, and the $ Q(r) $ profile relaxes to neutrality in a
distance of approximately a few $a$, thus suggesting once again that
a significant population of coions reside on the outer sides of
condensed counterions. Fluctuations of $ Q(r) $ for $ r \gg R_{0}
$ reflect density fluctuations which are much amplified because
of the volume factor $ 4 \pi r^{2} $. On the other hand, we
observe a nearly neutral region $ Q/|Q_{0} | \approx 1 $ extending
for the distance comparable to the Bjerrum length $ \ell_{B} $
outside the charge inversion layer.  Few ions exist in this
region. This shows establishment of enhanced order due to strong
Coulomb interactions.

The electrostatic potential drop across the charge distribution
peak corresponds to energy change $ e \Delta \varphi \approx 1.2
e^{2}/\epsilon a $, which is five times the thermal energy $
k_{B}T$. This implies strong binding of counterions to the
macroion and coions to the counterions.  In other words, this
manifests very strongly non-linear screening compared with
Debye-H\"uckel screening of weakly coupled cases. Of course, this
is by no means surprising given the small value of $\lambda$
Eq.(\ref{eq:lambda}), as mentioned above.

%\vspace{-.1cm}

Speaking about the dynamics of equilibration, it is interesting to
note that the buildup of counterions on the macroion occurs fairly
quickly, in about $ 100 \omega_{p}^{-1} $, which is of the order
of 100 picosec for the typical numerical values of parameters, as
suggested in Section \ref{sec:numbers}.  This time is much shorter
than overall relaxation time of the system, suggesting that
equilibration of plasma further away from macroion occurs fairly
slowly.  It is appealing to guess that this fast buildup of
screening (and even over-charging) layer is connected with the
fact of strongly non-linear correlated screening.

\subsubsection{Other regimes}

\begin{figure}
\centerline{\epsfxsize=0.4\textwidth \epsffile{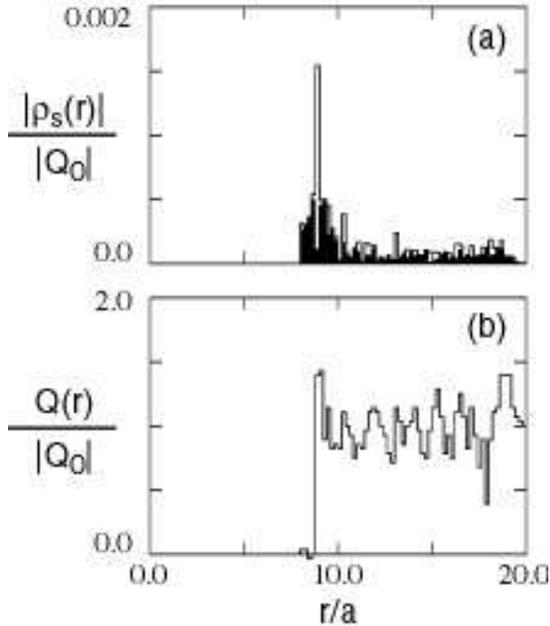} }

\vspace*{0.1cm}
\caption{Charge inversion of a "large macroion" - the system shown
in Fig.\protect\ref{standard_bird_eye}.  Plot format and notations
are the same as in Fig.\protect\ref{standard_plots}.  Simulation
parameters are also the same as those of
Fig.\protect\ref{standard_bird_eye}, except that the temperature is
adjusted to keep constant the quantity $ e^{2}/ \epsilon R_{0}  T
= const $.  Counterions are loosely bound to the macroion as $
Q_{0}  $ is fixed, and although the amount of inverted charge is
smaller compared to that in Fig.\protect\ref{standard_plots},
it is still significant. } \label{large_plots}
\end{figure}

\begin{figure} \centerline{\epsfxsize=0.4\textwidth
\epsffile{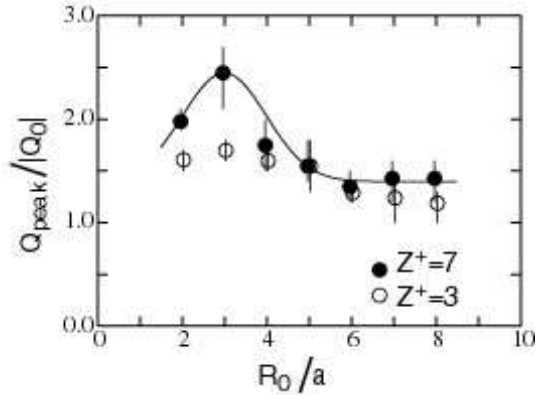} } \caption{Dependence of inverted charge on
the radius of macroion $ R_{0}  $ shown for the counterions with
the valence $ Z= 3 $ and 7. The charge of the macroion is $ Q_{0}
= -28e $, and the number of coions $ N^{-} = 335 $ (or 336)
corresponds to the density $ n^{-} \sim 1 \times 10^{-2} a^{-3} $.
The ordinate is the maximum of the integrated charge $ Q(r) $
(Eq.(\ref{eq:int-ch})), i.e. $ Q_{peak}= {\rm max}(Q(r)) $,
normalized by the macroion charge $ |Q_{0} | $. Each data point is
an average of three runs, and a vertical bar shows the range of
time variations.}\label{figure3}
\end{figure}

The charge inversion for the macroion with a large radius $ R_{0}
= 8a $ is depicted in Figs. \ref{large_bird_eye} and
\ref{large_plots}. Other parameters are the same as those of Fig.
\ref{standard_bird_eye}, except for the lower temperature ($
\Gamma_a = 78.4 $) to keep $ \Gamma_{Q} = const $
Eq.(\ref{eq:GammaQ}). We again observe sparsely distributed
counterions on the macroion surface. In this case, however,
binding of the counterions to the macroion is loose, and their
radial distribution in Fig. \ref{large_plots} (a) is almost as
broad as that of the coions. The counterion charge is better
canceled on each site by the condensed coions than in Fig.
\ref{standard_bird_eye}.

\vspace{-.3cm}

We note that the number of condensed ions to the macroion surface
in Fig. \ref{large_bird_eye} is $ N^{+} \sim 13 $ and $ N^{-} \sim
66 $, where the number of $ N^{+} $ is comparable to that in Fig.
\ref{standard_bird_eye}. This is consistent with the fact that
each counterion occupies, roughly, a neutralizing region on the
macroion surface, similar to the Wigner-Seitz cell of Wigner
crystal.  With charge density of the macroion surface $\sigma =
Q_{0} / 4 \pi R_{0} ^2$, the size of such neutralizing region, or
cell, is proportional to the size of the macroion: $e Z = \pi
\sigma R_{ws}^2$, or $ R_{ws}= 2R_{0}  (Ze/|Q_{0} |)^{1/2} $. In
other words, the neutralizing number of counterions $ (R_{0}
/R_{ws})^{2} $ stays unchanged as long as the macroion charge $
Q_{0}  $ is fixed. The inverted charge in Fig. \ref{large_plots}
(b) is about 40\% the original charge of the macroion, which is
less than that in Fig. \ref{standard_plots}. The electrostatic
potential drop across the macroion surface is consistently less
than the thermal energy, $ e \Delta \varphi \sim 0.05
e^{2}/\epsilon a <  k_{B}T \sim 0.09 e^{2}/\epsilon a $. The
linear Debye-H\"uckel theory nearly applies in this case.

We found similar features, based on identification of bound ions
in the peak of their radial distribution, also for the parameters
further away from our standard conditions.  For instance, we
mention here in passing the case of the counterions with smaller
valence $ Z= 3 $.  For them, it takes somewhat less than $ 1
\times 10^{3} \omega_{p}^{-1} $ to reach the stationary state, and
the attained peak height is lower, about 70\% the macroion charge,
as shown in Fig. \ref{figure6}. This will be discussed in greater
details in one of the sections below.

\vspace{-.3cm} 

\subsection{Changing macroion properties and temperature}

In the following figures, Figs. \ref{figure3}--\ref{figure7}, the
ordinate $ Q_{peak} $ is the maximum of the integrated charge of
the counterions plus coions, Eq.(\ref{eq:int-ch}). Each data point
is an average of three runs, and a vertical bar shows the range of
time variations and deviations among the runs.

The dependence of charge inversion on the radius of the macroion
is depicted in Fig. \ref{figure3}. For different values of the
radius, temperature is adjusted accordingly to keep unchanged the
value of $ \Gamma_{Q} \propto 1/(R_{0} T) $ Eq.(\ref{eq:GammaQ}). The
valence of the counterions is chosen either $ Z= 3$ or 7. The
number of counterions is $ N^{+}= 121 $ and $ N^{+}= 52 $ for $ Z=
3$ and 7, respectively, which is large compared to $ |Q_{0} |/Z e
$ required for charge neutralization of the macroion.   These
parameters are chosen in such a way that the number of coions,
which is determined by neutrality condition, is virtually fixed,
being $ N^{-}= 335 $ for the $ Z= 3$ case and $N^{-}=336$ for the
$ Z= 7$ case.  This corresponds to $r_s$ Eq.(\ref{eq:rs}) moderately
changing between $ 0.3a \sim 0.8a $, and $\lambda$
Eq.(\ref{eq:lambda}) changing between $ 0.02a \sim 2.8a $.

In Fig. \ref{figure3}, the inverted charge reaches its maximum for
the radius $ R_{0}  \approx 3a $ irrespectively of the valence $ Z
$. It falls off rapidly both for smaller and larger radii, and
becomes insensitive to the radius of the macroion for $ R_{0} /a
\gg 1 $. The net amount of the inverted charge is about 70\% of
the bare macroion charge $ Q_{0}  $ for $ Z= 3 $; it increases up
to 150\% of $ Q_{0}  $ for $ Z= 7 $.
We find that the charge inversion reaches maximum also at virtually
the same radius $ R_{0}  \approx 3a $ even for the smaller number of
counterions $ N^{+}=15 $ ($ Z= 7 $), or for larger macroion charge
$ Q_{0} = -42e $.

It is not difficult to understand qualitatively why the charge
inversion decreases at both small and large values of macroion
radius $ R_{0}  $, reaching a maximum in between. When $R_{0} $
gets very large, the lateral spacings between bound counterions
become too long to maintain correlations between them; on the
other hand, when $R_{0} $ gets too small, the increased repulsion
of the inverted charge from the macroion becomes dominant.

%\vspace{-.3cm}
\begin{figure}
\centerline{\epsfxsize=0.4\textwidth \epsffile{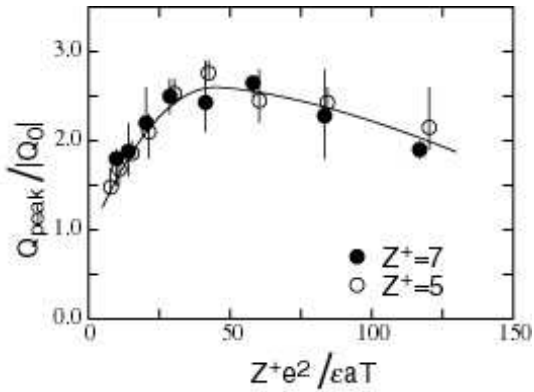} }

\caption{Dependence of inverted charge on
temperature shown for counterions with different valence $ Z $,
which is given by a master curve. The abscissa is the ratio of the
Coulomb energy of the counterions to thermal energy, $ Z
e^{2}/\epsilon a T $. The radius of macroion is $ R_{0} = 3a $,
and the number of coions is kept nearly the same, $ N^{-} \sim 335
$ or 336 for $ Z= 5 $ and 7.} \label{figure4}
\end{figure}

The effect of temperature on charge inversion is shown in Fig.
\ref{figure4}. In this figure, the abscissa is $ \Gamma_a= Z
e^{2}/\epsilon aT $ Eq.(\ref{eq:Gamma}).  As the figure indicates,
the inverted charge for different values of valence form a
master curve when plotted against $ \Gamma_a $.  The charge
inversion is maximized at the intermediate temperature
corresponding to $ \Gamma_a \sim 45 $, or $ Z e^{2}/\epsilon R_{0}
T \sim 15 $ ($ R_{0} = 3a $). The value of the Debye length is $
r_{s} \approx 0.6a $ for $ Z=3 $ and $ r_{s} \approx a $ for $ Z=7
$. For the low temperature side, $ \Gamma_a \sim 100 $, the
integrated charge distribution $ Q(r) $ is sharply peaked as that
of Fig. \ref{standard_plots} (b), while at the high temperature
side, $ \Gamma_a \sim 10 $, this distribution $Q(r)$ is rugged and
fluctuates considerably with time.
It is consistent with that the maximal charge inversion
is achieved through competition of counterion attachment to the
macroion and coion condensation on the counterions.  Lower
temperatures are favored for the former due to larger Coulomb
binding energy, and higher temperatures are better to suppress the
latter due to enhanced thermal motion.

Figure \ref{figure5} shows that charge inversion $ Q_{peak}/|Q_{0}
| $ is insensitive to the charge content of the macroion $ Q_{0} $
for fixed value of $ \Gamma_{Q}= Q_{0} e/\epsilon R_{0} T $ ($
\Gamma_a= 4.2Z$ or $6Z$). The number of counterions attached to
the macroion surface is in the range $8 \sim 15$ for $ |Q_{0} |=
(14 \sim 42) e$ and $ Z= 7 $, which is a few times that of the
neutralizing number of counterions, $ |Q_{0} |/Ze $.

\vspace{-.3cm} \begin{figure} \centerline{\epsfxsize=0.4\textwidth
\epsffile{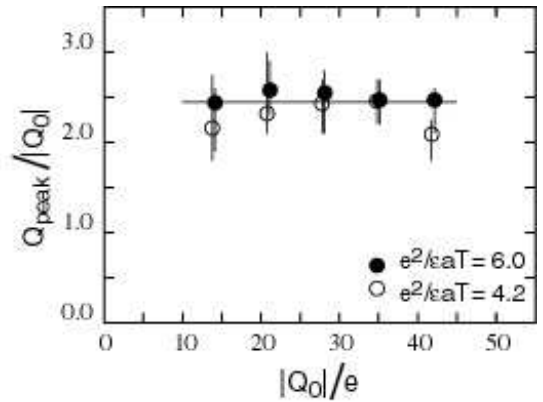} } \caption{Dependence of inverted charge on
the macroion charge $ Q_{0}  $.  The radius of macroion is $ R_{0}
= 3a $, the valence of counterions is $ Z= 7 $. Temperature is
adjusted to keep the binding energy $ \Gamma_{Q}= |Q_{0} |
e/\epsilon R_{0}  T $ constant as $ Q_{0}  $ varies.}
\label{figure5}
\end{figure}

We note in passing that the geometrical capacity of the surface,
controlled by non-Coulomb short range forces is still very far
from exhausted, $ 4 \pi R_{0} ^{2}/ \pi a^{2} \sim 36 $.  The
regime of closed and almost closed packing of the bound spheres on
the macroion is examined in the recent work\cite{NSpriv}.
Interestingly, the {\it effective} valence of the counterions $
Z_{eff} $, which is the charge of the counterion minus that of the
condensed ions, increases with the charge of the macroion; it is $
Z_{eff} \sim 0.25 Z $ for $ Q_{0} = -14e $ and is $ Z_{eff} \sim
0.4 Z $ for $ Q_{0} = -42e $.

\subsection{Changing counterion properties}

\begin{figure} \centerline{\epsfxsize=0.4\textwidth
\epsffile{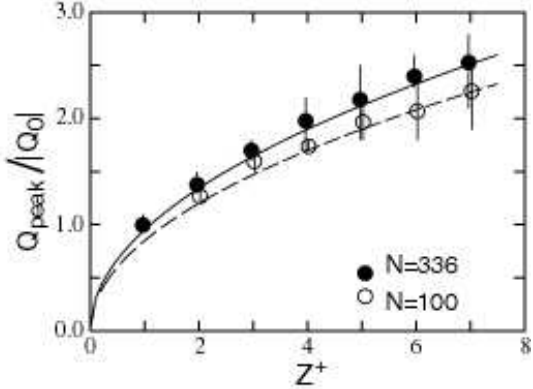} } \caption{Amount of inverted charge $
Q_{peak} $ increasing monotonically with the valence of
counterions $ Z $. The inverted charge is well fit by $ Q_{peak}
\sim Z^{1/2} $ for $ Z \le 5 $, and $ Q_{peak} \sim Z $ for $ Z
\ge 5 $.  Note that charge inversion occurs only for multivalent
counterions, i.e. $ Z \ge 2 $.} \label{figure6}
\end{figure}

The dependence of inverted charge on the valence of the
counterions is depicted in Fig. \ref{figure6}. Here, the macroion
charge and radius are $ Q_{0} = -28e $ and $ R_{0} = 3a $,
respectively, and temperature is fixed such that $ \Gamma_a /Z =
e^{2}/\epsilon aT= 4.2 $. It is emphasized that no charge
inversion is observed for monovalent counterions. The amount of
the inverted charge $ Q_{peak} $ increases with the valence, which
is well scaled by $ Q_{peak} \sim Z^{1/2} $ for $ Z \le 5 $. The
$ Z \ge 5 $ part can be fit by $ Q_{peak} \sim Z $. The inverted
charge is also an increasing function of the number of counterions
and coions, as seen by the difference of the two curves for two
densities in the figure.

\vspace{-.3cm} 

%\vspace{-.3cm}
\begin{figure}
\centerline{\epsfxsize=0.4\textwidth \epsffile{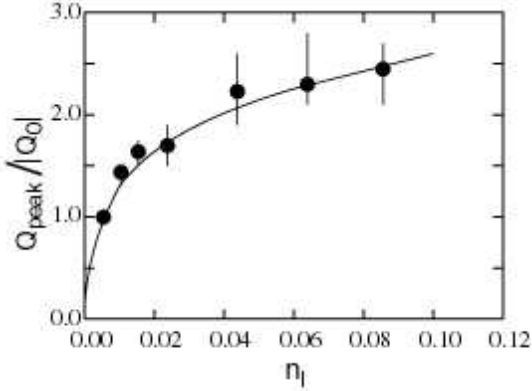} }
\caption{Dependence of inverted charge on the ionic strength $
n_{I} = (Z^{2} N^{+} + N^{-})/V $, where $V$ is the volume of the
simulation cell.
% $ V= 4 \pi (R_{M}^{3} -R_{0} ^{3})/3 $.
Charge neutrality of the system is maintained. The guide curve is
$ n_{I}^{1/2} $ for $ n_{I} < 0.02/a^{3} $, and $ n_{I} $ for $
n_{I} > 0.05/a^{3} $. The macroion radius is $ R_{0} = 3a $,
charge $ Q_{0} = -28e $, the valence of counterions $ Z= 7 $, and
$ \Gamma_a = 29.4 $.} \label{figure7}
\end{figure}

The dependence of inverted charge on the ionic strength, $ n_{I}=
(Z^{2} N^{+} +N^{-})/ V $, is shown in Fig. \ref{figure7}, where $
V= 4 \pi (R_{M}^{3} -R_{0} ^{3})/3 $ is the domain volume. The
amount of inverted charge $ Q_{peak}/|Q_{0} | $ increases
monotonically with the ionic strength. The functional form of the
scaling changes at $ n_{I} \sim 0.05/a^{3} $, as shown by fitting
curves. The ionic strength of a $ {\rm Ca}^{2+} $ ion and
neutralizing coions in every $ 10 \AA $ cube yields $ 0.048/a^{3}
$ for $ a= 2 \AA $. The scaling $ Q_{peak} \sim n_{I}^{1/2} $ for
the low ionic strength $ n_{I} < 0.02/a^{3} $ smoothly joins a
linear scaling $ Q_{peak} \sim n_{I} $ for high ionic strength $
n_{I} > 0.05/a^{3} $. The non-dimensional parameter of the
theory\cite{NguGS}, $ \zeta= (R_{ws}/r_{s})^{2} = 12a \Gamma_a
N_{ci} (e/|Q_{0} |)(R_{0} ^{2}/R_{M}^{3}) $, is calculated to be
0.7 for $ n_{I} \sim 0.01/a^{3} $ and $ Z= 7 $. The theory expects
$ Q^{(th)} \sim (N_{ci} Z)^{1/2} $ for $ \zeta \ll 1 $, and $
Q^{(th)} \sim N_{ci} Z $ for $ \zeta \gg 1 $. The present
simulation results agree with this theoretical prediction.

\vspace{-.3cm} \begin{figure} \centerline{\epsfxsize=0.4\textwidth
\epsffile{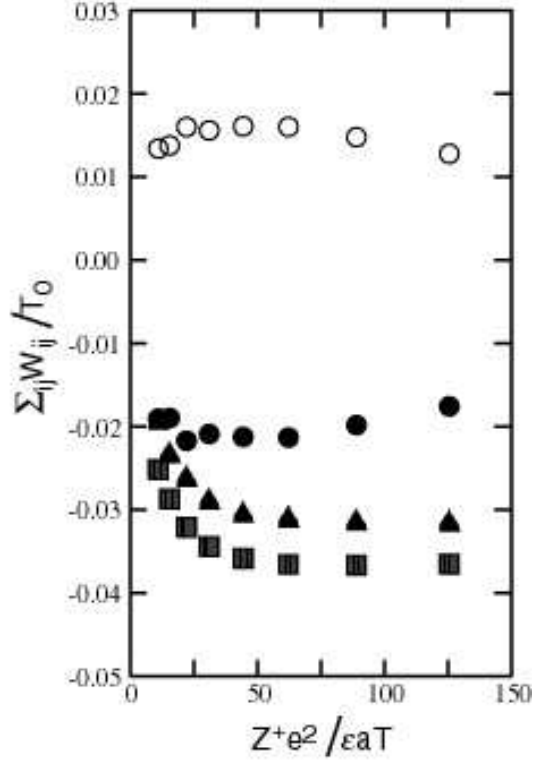} } \caption{Potential energy shown as a
function of $ Ze^{2}/\epsilon aT $ for the runs with $ Z= 7 $ (cf.
Fig. \protect\ref{figure4}).  The filled and open circles
correspond to the potential energy of interaction of a macroion
with counterions and coions, respectively, the triangles those
between counterions and coions, and the squares the total
potential energy.} \label{figure8}
\end{figure}

\subsection{Measuring potential energy}

The potential energy presented in Fig. \ref{figure8} is in line
with the tendency of charge inversion dependence on variations of
the Coulomb coupling parameter (cf. Fig. \ref{figure4}).  The
potential energy for the interactions between counterions and the
macroion (solid circles) is negative (attractive) and is minimized
at the intermediate value of $ \Gamma_{a}= Z e^{2}/\epsilon aT
\sim 50 $ where
largest charge inversion is obtained. The potential energy of
interactions between the counterions and coions, depicted by
triangles, decreases remarkably with the increase in $ \Gamma_{a} $.
This corresponds to massive condensation of
coions onto the counterions (similar to Manning-Onsager
condensation) at low temperatures. This reduces the effective
valence of the counterions, and the binding of counterions on the
macroion surface is weakened, which tends to suppress the charge
inversion.  Thus, charge inversion becomes largest at the
intermediate value of $ Z e^{2}/\epsilon aT $, as stated above.
The potential energy of interaction between coions and the
macroion (open circles) is positive (repulsive), and is maximized
where the coions are closely located with the macroion by
condensation to the counterions, namely at $ \Gamma_{a} \sim 50 $.
On the other hand, the total potential energy (squares)
decreases with the increase in the coupling parameter $ \Gamma_{a} $.

\section{Conclusion}

In this paper, we showed the occurrence of giant charge inversion
and examined its parameter dependences with the use of molecular
dynamics simulations.  The charge inversion was found to be based
on the strong correlations of the multivalent counterions and
coions, particularly on the surface of the macroion. Specifically,
charge inversion was observed under the conditions for which the
Coulomb coupling parameter was significantly larger than unity, $
\Gamma  \gg 1 $.  At the same time, charge inversion occurred only
in the presence of multivalent counterions with $ Z \ge 2 $. The
counterions were attached to the surface of the macroion, while
monovalent coions tended to condense on the counterions. This
condensation, similar to Bjerrum pairing, is therefore identified
as the factor limiting the amount of charge inversion. The amount
of the inverted charge $ Q_{peak} $ was maximal at rather small
radius of the macroion, and leveled off when radius becomes
larger.  It scaled linearly with the charge of the macroion $
Q_{0}  $, and the ratio $ Q_{peak}/|Q_{0} | $ was independent of
the macroion charge.

With respect to the valence $ Z $ and the ionic strength $ n_{I}=
(Z^{2} N^{+} +N^{-})/V $, the amount of inverted charge scaled as
$ Q \sim (Z n_{I})^{1/2} $ for the valence $ Z \le 5 $ or $ n_{I}
\le 0.02/a^{3} $. As noted in Sec.III C, this ionic strength
corresponds to a $ {\rm Ca}^{2+} $ ion in every $ 10 \AA $ cube.
The inverted charge scaled as $ Q \sim Z n_{I} $ for $ Z > 5 $ or
$ n_{I} > 0.05/a^{3} $.  This agreed with the theory of giant
charge inversion\cite{NguGS}. The net inverted charge of nearly up
to 160\% the bare charge of the macroion was achieved at the
medium temperature $ Z e^{2}/\epsilon R_{0}  T \sim 15 $, due to
the competition of multivalent counterion attachment to the
macroion and monovalent coion condensation on the counterions; the
former was stronger at lower temperatures, and the latter was
suppressed at higher temperatures.

In the present study, the macroion was assumed to be immobile.
From the application points of view, it might be informative to
study the distribution of counterions and coions around a moving
macroion and also the effect of an applied electric field.  The
study of such cases is reported in a separate paper. The results
indicate that a formed complex of a macroion and counterions
drifts along the electric field in the direction implied by the
inverted charge, and that charge inversion is not altered until
the electric field exceeds a critical value \cite{Tan}.

\acknowledgments

The authors are highly grateful to Professor Boris I. Shklovskii
whose suggestions helped them to perform the current study. We
also thank him for reading the paper and giving them valuable
comments.

\end{document}